# Disorder effect in low dimensional superconductors

T Xiang and J M Wheatley

*I.R.C. in Superconductivity, University of Cambridge, Madingley Road, Cambridge CB3 0HE, United Kingdom*

The quasiparticle density of states (DOS), the energy gap, the superfluid density $\rho_s$, and the localization effect in the s- and d-wave superconductors with non-magnetic impurity in two dimensions (2D) are studied numerically. For strong (unitary) scatters, we find that it is the range of the scattering potential rather than the symmetry of the superconducting pairing which is more important in explaining the impurity dependences of the specific heat and the superconducting transition temperature in Zn doped YBCO. The localization length is longer in the d-wave superconducting state than in the normal state, even in the vicinity of the Fermi energy.

In this paper we present numerical results for the density of states and other physical quantities in superconducting states with strong potential scattering from non-magnetic impurities in Zn doped YBCO. A comparison with experiments as well as other theoretical results is also given.

The model we study is the BCS mean-field Hamiltonian with disorder which is defined by

$$H = -t \sum_{<\mathbf{rr'}>\sigma} c^\dagger_{\mathbf{r}\sigma} c_{\mathbf{r'}\sigma} + \sum_{\mathbf{r}\tau}(\Delta_{\mathbf{r}\tau} c^\dagger_{\mathbf{r}\uparrow} c^\dagger_{\mathbf{r}+\tau\downarrow} + h.c.)$$
$$+ \sum_{\mathbf{r}\sigma}\Big(\sum_{\mathbf{r}_{imp}} V(\mathbf{r}-\mathbf{r}_{imp}) - \mu\Big) c^\dagger_{\mathbf{r}\sigma} c_{\mathbf{r}\sigma}, \quad (1)$$

where $<\mathbf{rr'}>$ denotes nearest neighbors, $\mu$ is the chemical potential, and $\mathbf{r}_{imp}$ are the positions of random impurities. We assume the random potential $V(\mathbf{r}-\mathbf{r}_{imp}) = V_0 \delta_{\mathbf{r},\mathbf{r}_{imp}} + V_1 \delta_{<\mathbf{rr}_{imp}>}$. When $V_1 = 0$, it is a $\delta$-function potential, otherwise it is finite ranged [1]. $\Delta_{\mathbf{r}\tau}$ is the superconducting order parameter. It should be determined self-consistently from the relation $\Delta_{\mathbf{r}\tau} = J <c_{\mathbf{r}\uparrow} c_{\mathbf{r}+\tau\downarrow}>$, where J (assumed disorder independent) is the coupling constant. In the absence of disorder, $\Delta_{\mathbf{r}\tau}$ is translation invariant with a symmetry compatible with the point group of the lattice. In this paper only the on-site s-wave pairing symmetry, $\Delta_{\mathbf{r}\tau} = \Delta \delta_{\tau,0}$, and the d-wave pairing symmetry, $\Delta_{\mathbf{r}\tau} = \Delta(\delta_{\tau,\pm\hat{x}} - \delta_{\tau,\pm\hat{y}})$, will be considered.

In the presence of disorder, $\Delta_{\mathbf{r}\tau}$ is no longer translationally invariant. However, if the fluctuation of $\Delta_{\mathbf{r}\tau}$ in space is not too strong, the symmetry of the gap function is preserved on a sufficiently large scale. To simplify the calculation, we approximate $\Delta_{\mathbf{r}\tau}$ in (1) by their average in space, $\bar{\Delta}_\tau = \sum_{\mathbf{r}} \Delta_{\mathbf{r}\tau}/N$ (N the lattice size), so that only $\bar{\Delta}_\tau$ is determined self-consistently. The results obtained from this approximation can be taken as a qualitative guide to the properties of superconductors in the limit of very strong disorder, but are accurate for dilute impurities.

Fig. 1 shows the DOS [2,3] for the s- and d-wave superconductors with strong impurity potentials for various impurity concentrations x obtained using a recursion method. For both symmetries the DOS near zero energy increases with x and becomes comparable with the average DOS at x∼0.07 in the finite range model. This behavior agrees qualitatively with the variation of $\gamma(0)$ (linear specific heat coefficients) with Zn concentrations in the Zn doped YBCO [4]. For the d-wave superconductor with a strong $\delta$-function potential, a small DOS peak is observed near the zero energy in consistent with the resonant scattering theory of the d-wave superconductor in the unitary scattering limit [5]. As the impurity potential becomes small, this peak disappears, but the DOS at the zero energy is always found to be finite. We have not found evidence for the singularity at the zero energy which was predicted recently by Nersesyan et al [6].

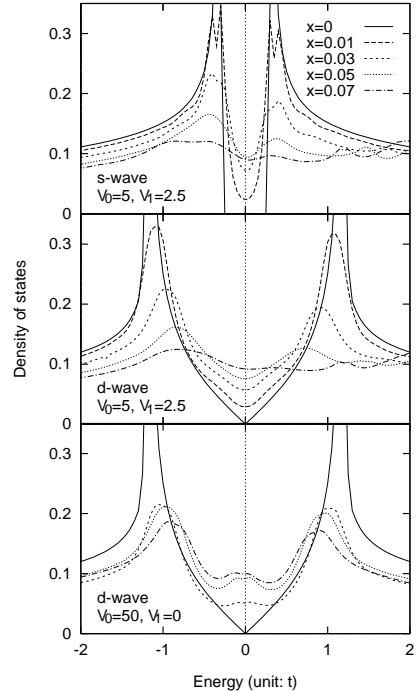

FIG. 1. DOS as a function of energy for the s- and d-wave superconductors at half-filling.

In Figure 2 we compare the energy gap and superfluid density $\rho_s$ as functions of x for the s- and d-wave superconductors at zero temperature [3]. $\rho_s$ is obtained by directly calculating the current-current correlator in the ground state. A finite range impurity potential suppresses both $\bar{\Delta}$ and $\rho_s$ more strongly than a $\delta$-function potential. Comparing the result with the drop of the superconducting transition temperature with x in the Zn doped YBCO [7], we find that our result for both the s- and d-wave superconductors with a finite range impurity potential can give a qualitative fit with the experiment, while the result for the s- and d-wave superconductors with the $\delta$-function potential decreases much too slowly to fit the experiment. This is in agreement with our estimate for the scattering phase shift from the resistivity measurements [1].

We have also examined the localization effect in the d-wave superconductor by calculating the localization length of the quasiparticle using the scaling method. We find that the localization length in a superconducting state is far longer than that in a corresponding non-superconducting state for all energies, hence the localization effect is much weaker in a superconducting state than that in a normal state, and is unlikely to influence the behavior of $\rho_s$ for example. The scaling function for the d-wave superconducting state is the same as for the corresponding non-superconducting state consistent with a single parameter scaling theory.

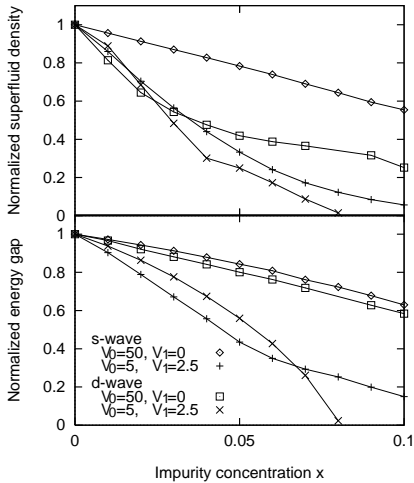

FIG. 2. Normalized energy gap and superfluid density vs x for the s- and d-wave superconductors at half filling on a 10×10 lattice at zero temperature.